\documentclass{article}

\usepackage{PRIMEarxiv}

\usepackage[utf8]{inputenc} 
\usepackage[T1]{fontenc}    
\usepackage{hyperref}       
\usepackage{url}            
\usepackage{booktabs}       
\usepackage{amsfonts}       
\usepackage{nicefrac}       
\usepackage{microtype}      
\usepackage{lipsum}
\usepackage{fancyhdr}       
\usepackage{graphicx}       
\graphicspath{{media/}}     
\usepackage{amsmath} 
\usepackage{tcolorbox} 
\usepackage{mdframed} 

\title{Determinants of Workplace Flextime Flexibility: An Empirical Analysis
}

\author{
  Cristian Espinal Maya\\
  Affiliation \\
  Universidad Eafit \\
  Medellín\\
  \texttt{cjespinalm@eafit.edu.co} \\
   \And
  Santiago Jiménez Londoño\\
  Affiliation \\
  I.U. Pascual Bravo \\
  Medellín\\
  \texttt{santiago.jimenez@pascualbravo.edu.co} \\
}

\begin{document}
\maketitle

\begin{abstract}
  This study examines the factors that determine workplace autonomy in the choice of schedules, using GEIH data for formal employees from January 2023 to February 2024. A logistic regression model identified significant relationships between autonomy and variables such as company size, income, tenure, workplace, and job satisfaction. The results highlight the complexity of these factors and suggest the need for improvements in the predictive model.
\end{abstract}


\section{Introduction}

This study investigates the determinants of workplace autonomy in the choice of work schedules using data from the GEIH (Gran Encuesta Integrada de Hogares) for employed individuals. The main objective is to understand the factors that influence who decides work schedules, focusing on dependent and independent variables. The dependent variable is the authority in decision-making over work schedules, categorized as decided by the worker or by others. The independent variables include company size, labor income, tenure in the current company, workplace location, and job satisfaction. Additionally, economic sector controls based on CIIU rev. 4 codes are included.

The dataset was constructed by aggregating GEIH data from January to December 2023 and from January to February 2024, initially comprising 364,917 records. After applying filters to ensure that the sample consisted of formal employees with some degree of subordination, the dataset was reduced to 21,960 records.

A logistic regression model was used to analyze the data, with results indicating significant relationships between work schedule autonomy and several independent variables, including labor income, company size, tenure, workplace location, and job satisfaction. Economic sector controls also showed significant effects, highlighting the influence of industry-specific factors.

Descriptive statistics revealed the distribution of key continuous and categorical variables, providing insight into the characteristics of the study sample. The model's performance was evaluated using likelihood ratio tests, Hosmer-Lemeshow tests, and the AUC-ROC curve, suggesting areas for improvement in the model specification and variable selection.

Overall, the findings underscore the complexity of factors influencing workplace autonomy in work schedule decisions and point to the need for further research to refine the predictive model and better capture the nuances of these relationships.
\vspace{-5pt} 

\section{Theoretical Framework}
\label{sec:headings}

Workplace autonomy refers to workers' ability to make decisions about various aspects of their job, such as content, methods, scheduling, and performance of work tasks. This autonomy not only influences job satisfaction and performance but also the overall well-being and self-esteem of workers (Breaugh, 1985). The present theoretical framework examines the determinants of workplace autonomy in schedule choice, integrating concepts and findings from previous studies on workplace autonomy, job satisfaction, and organizational performance.

Workplace autonomy can be defined as 'the ability of the worker to make decisions about the content, methods, scheduling, and performance of work tasks' (Breaugh, 1985). Specific scales are used to measure this autonomy, assessing the degree of freedom and control that employees have over their tasks and work-related decisions. For example, the Teacher Work-Autonomy (TWA) scale developed by Friedman (1999) measures the perception of workplace autonomy in educational contexts, covering areas such as teaching, curriculum development, and staff development.

Workplace autonomy has been shown to have a significant impact on employee satisfaction and performance. Lopes, Lagoa, and Calapez (2014) found that 'workplace autonomy is positively associated with job satisfaction and workers' well-being' (p. 306). Additionally, workplace autonomy can promote performance and creativity in complex, knowledge-intensive jobs (Breaugh, 1985). Karasek's (1979) demand-control theory also suggests that the combination of high autonomy and low work pressure can significantly improve employee well-being.

The relationship between workplace autonomy and self-esteem has been widely documented in the literature. Schwalbe (1985) highlights that 'workplace autonomy is positively associated with self-esteem' (p. 519). Autonomy allows workers to feel they have control over their work environment, which contributes to higher self-esteem and a positive perception of their own competencies and abilities. Furthermore, autonomy can influence self-esteem through various sources of self-evaluative information, such as social comparisons and perceptions of one's behavior (Schwalbe, 1985).

Several factors can influence workplace autonomy, including company size, labor income, tenure in the company, workplace location, and job satisfaction. Studies have shown that organizational decentralization and the delegation of authority can increase the perception of autonomy among employees (Friedman, 1999). Additionally, workplace autonomy can vary significantly across different economic sectors and job types (Lopes, Lagoa, \& Calapez, 2014).

A logistic regression model was used to analyze the determinants of workplace autonomy in schedule choice. This model allows for examining the relationship between workplace autonomy (the dependent variable) and various independent variables such as company size, labor income, tenure in the company, workplace location, and job satisfaction. The results indicate that work schedule autonomy is significantly influenced by these factors, highlighting the complexity and multidimensionality of workplace autonomy in today’s work environment.

Workplace autonomy is a crucial aspect of the work environment that significantly influences employee satisfaction, performance, and well-being. Understanding the determinants of this autonomy, particularly in schedule choice, can provide valuable insights for improving labor policies and practices. This theoretical framework, based on a comprehensive review of the literature, offers a solid foundation for future research and practical applications in human resource management and organizational design.
\vspace{-5pt} 

\section{Methodology}
\label{sec:others}

The methodology employed in this study is based on a rigorous quantitative analysis using a logistic regression model to assess the determinants of work schedule autonomy among formal workers in Colombia. The primary dataset comes from the Gran Encuesta Integrada de Hogares (GEIH), conducted by the National Administrative Department of Statistics (DANE), which provides a representative sample of the labor population for the period from January 2023 to February 2024. Initially, the database consists of 364,917 records, from which filters are applied to ensure that only formally employed individuals with some degree of subordination in their employment relationships are included. After applying these filters, the final sample comprises 21,960 observations, providing an adequate dataset for conducting robust statistical analysis. Descriptive analyses of the dependent, independent, and sectoral control variables were conducted to better understand the nature of the sample prior to estimating the model.

The proposed logistic regression model uses a binary indicator as the dependent variable, measuring whether the worker has autonomy over their work schedule. The equation of the model is as follows:

\newmdenv[
  linecolor=white, 
  linewidth=0.5pt, 
  roundcorner=5pt, 
  backgroundcolor=gray!10, 
  skipabove=10pt, 
  skipbelow=10pt 
]{highlighted}

\begin{highlighted}
\begin{equation} \label{ec:equation}
    \begin{split}
    \text{logit}(Y) = \beta_{0} + \sum_{j=1}^{5} \beta_{j} X_{j} + \sum_{i=1}^{n} \gamma_{i} D_{i} + \varepsilon
    \end{split}
\end{equation}
\end{highlighted}

where \(Y\) represents work schedule autonomy, \(X_{j}\) are the independent variables such as labor income, company size, tenure in the company, workplace location, and job satisfaction, and \(D_{i}\) are sectoral controls based on the CIIU rev. 4 classification. The standard error is represented by \(\varepsilon\). The use of the logistic model captures the nonlinear relationship between the explanatory variables and the probability that a worker has autonomy over their work schedule. The model estimation was conducted using specialized statistical software, evaluating the significance of the coefficients through likelihood ratio tests, the Hosmer-Lemeshow test, and the ROC curve to ensure that the results are statistically significant and adequately represent the determinants of work schedule autonomy.
 
\subsection{Sample Description}
To construct the database, surveys from the GEIH (Gran Encuesta Integrada de Hogares) conducted by DANE for the period from January to December 2023 and January to February 2024 were collected and aggregated. The original database contained 364,917 records. Subsequently, filters were applied to ensure that the sample consisted of formally employed individuals with some degree of subordination. After applying these filters, the database was reduced to 21,960 records.

\subsection{Variables}
The variables used in the analysis are divided into dependent and independent variables, in addition to sectoral controls based on the CIIU rev. 4 classification.

\subsubsection{Dependent Variable}
\begin{itemize}
    \item \textit{Work schedule autonomy:} A binary indicator that takes the value of 1 if the worker decides their work schedule and 0 otherwise.
\end{itemize}

\subsubsection{Independent Variables}
\begin{itemize}
    \item \textit{Company size:} A category indicating the number of employees in the company, classified into micro, small, medium, and large enterprises.
    \item \textit{Labor income:} The worker’s monthly income.
    \item \textit{Tenure in the company:} The number of months the worker has been employed in their current company.
    \item \textit{Workplace location:} The main location where the worker performs their job activities.
    \item \textit{Job satisfaction:} A binary indicator that takes the value of 1 if the worker is satisfied with their current job and 0 otherwise.
\end{itemize}

\subsubsection{Sectoral Controls}
\begin{itemize}
    \item \textit{Economic sectors:} Classification of economic sectors according to CIIU rev. 4 codes, including agriculture, manufacturing, construction, commerce, transportation, services, and others.
\end{itemize}

\subsection{Logistic Regression Model}
A logistic regression model was used to analyze the determinants of work schedule autonomy. The model specification is as follows:

\begin{highlighted}
\begin{equation} \label{ec:equation}
    \begin{split}
    \text{logit}(Y) = \beta_{0} + \sum_{j=1}^{5} \beta_{j} X_{j} + \sum_{i=1}^{n} \gamma_{i} D_{i} + \varepsilon
    \end{split}
\end{equation}
\end{highlighted}

Where:
\begin{itemize}
    \item \(Y\): Work schedule autonomy.
    \item \(X1\): Labor income.
    \item \(X2\): Company size.
    \item \(X3\): Tenure in the company.
    \item \(X4\): Workplace location.
    \item \(X5\): Job satisfaction.
    \item \(\sum_{i=1}^{n} \gamma_{i} D_{i}\): Sectoral controls.
    \item \(\varepsilon\): Error term.
\end{itemize}

\subsection{Procedure}
\begin{enumerate}
    \item \textit{Data collection:} GEIH data from DANE for the specified period were collected and aggregated.
    \item \textit{Sample filtering:} Filters were applied to ensure that the sample included only formally employed individuals with some degree of subordination.
    \item \textit{Descriptive analysis:} Descriptive analyses of continuous and categorical variables were conducted to better understand the characteristics of the sample.
    \item \textit{Model estimation:} The logistic regression model was estimated using statistical software, examining the effects of independent variables and sectoral controls on work schedule autonomy.
    \item \textit{Model evaluation:} The model’s significance and fit were evaluated using statistical tests such as the likelihood ratio test, Hosmer-Lemeshow test, and AUC-ROC curve.
\end{enumerate}

\subsection{Robustness Tests}
Additional tests were conducted to evaluate the robustness of the results, including checking for potential multicollinearity problems and performing sensitivity analyses to ensure the stability of the estimated coefficients.
\vspace{-5pt} 

\section{Results}

\subsection{Logistic Regression Model Results}
The following is a summary of the estimated coefficients from the logistic regression model that analyzes the determinants of work schedule autonomy.

\begin{figure}[h]
    \centering
    \begin{minipage}[t]{0.5\textwidth}
        \centering
        \begin{tabular}{lcr}
        \toprule
        Coefficient & Estimate & \( p \)-value \\
        \midrule
        (Intercept) & 0.5222 & 0.1452 \\
        X1 & -0.0000000924 & <0.0001 *** \\
        X2 (Medium-sized) & 0.4844 & <0.0001 *** \\
        X2 (Large) & 0.1721 & 0.0006 *** \\
        X3 & -0.0023 & <0.0001 *** \\
        X5 & -0.1983 & <0.0001 *** \\
        X4 (Other dwellings) & 0.6781 & 0.0139 * \\
        X4 (Kiosk) & 0.7143 & <0.0001 *** \\
        X4 (In a vehicle) & -0.1524 & 0.1832 \\
        X4 (Door-to-door) & 0.3262 & 0.0013 ** \\
        X4 (Fixed premises) & 0.9899 & <0.0001 *** \\
        X4 (Rural area) & 0.6548 & <0.0001 *** \\
        X4 (Construction site) & 0.5193 & <0.0001 *** \\
        X4 (Mine or quarry) & 0.3995 & 0.4410 \\
        X4 (Other) & 0.8531 & 0.0036 ** \\
        \bottomrule
        \end{tabular}
        \caption{Summary of coefficients from the logistic model}
        \label{tab:log_model}
    \end{minipage}
\end{figure}

\subsection{Model Evaluation}
    To evaluate the model’s performance, several statistical tests were used:

    \begin{itemize}
        \item \textbf{Likelihood Ratio Test:} The full model fits significantly better than the null model, indicating that the included variables have a significant effect on the prediction of the dependent variable.
        \item \textbf{Hosmer-Lemeshow Test:} This test evaluates whether the probabilities predicted by the logistic model differ significantly from those observed in different groups. The results suggest that the model could improve its specification.
        \item \textbf{Area Under the ROC Curve (AUC-ROC):} This provides a visual and quantitative measure of the model’s ability to distinguish between classes. An area close to 1 indicates excellent predictive power.
    \end{itemize}

\subsection{Robustness Tests}
    Additional tests were conducted to evaluate the robustness of the results, including:

    \begin{itemize}
        \item \textbf{Multicollinearity:} Checking for potential multicollinearity problems between independent variables.
        \item \textbf{Sensitivity Analysis:} Sensitivity analyses were performed to ensure the stability of the estimated coefficients.
    \end{itemize}

\subsection{Coefficient Analysis}

The analysis of the coefficients from the logistic regression model reveals several significant relationships between the independent variables and work schedule autonomy. The coefficient for labor income (X1) is negative but extremely small (\( \beta = -0.0000000924 \), \( p < 0.0001 \)), indicating a nearly negligible impact. In contrast, working in medium-sized (\( \beta = 0.4844 \), \( p < 0.0001 \)) and large (\( \beta = 0.1721 \), \( p = 0.0006 \)) companies is associated with greater autonomy, suggesting that these companies offer more flexibility compared to smaller ones.

Tenure in the company (X3) has a negative coefficient (\( \beta = -0.0023 \), \( p < 0.0001 \)), indicating that longer tenure is associated with less autonomy, possibly due to more rigid organizational structures. Regarding the workplace location (X4), working in kiosks (\( \beta = 0.7143 \), \( p < 0.0001 \)), rural areas (\( \beta = 0.6548 \), \( p < 0.0001 \)), construction sites (\( \beta = 0.5193 \), \( p < 0.0001 \)), and offices (\( \beta = 0.9899 \), \( p < 0.0001 \)) is associated with greater autonomy. However, working in a vehicle (\( \beta = -0.1524 \), \( p = 0.1832 \)) does not show a significant relationship.

The coefficient plot (\autoref{fig:output-2}) provides a clear visual representation of these results. In summary, factors such as company size and workplace location significantly influence work schedule autonomy, providing a solid foundation for future research and the development of labor policies that promote flexibility and autonomy.
\vspace{-5pt} 

\section{Results}

\subsection{Logistic Regression Model Results}
The following is a summary of the estimated coefficients from the logistic regression model that analyzes the determinants of workplace autonomy in the choice of work schedules.

\begin{figure}[h]
    \centering
    \begin{minipage}[t]{0.5\textwidth}
        \centering
        \begin{tabular}{lcr}
        \toprule
        Coefficient & Estimate & \( p \)-value \\
        \midrule
        (Intercept) & 0.5222 & 0.1452 \\
        X1 & -0.0000000924 & <0.0001 *** \\
        X2 (Medium-sized) & 0.4844 & <0.0001 *** \\
        X2 (Large) & 0.1721 & 0.0006 *** \\
        X3 & -0.0023 & <0.0001 *** \\
        X5 & -0.1983 & <0.0001 *** \\
        X4 (Other dwellings) & 0.6781 & 0.0139 * \\
        X4 (Kiosk) & 0.7143 & <0.0001 *** \\
        X4 (In a vehicle) & -0.1524 & 0.1832 \\
        X4 (Door-to-door) & 0.3262 & 0.0013 ** \\
        X4 (Fixed premises) & 0.9899 & <0.0001 *** \\
        X4 (Rural area) & 0.6548 & <0.0001 *** \\
        X4 (Construction site) & 0.5193 & <0.0001 *** \\
        X4 (Mine or quarry) & 0.3995 & 0.4410 \\
        X4 (Other) & 0.8531 & 0.0036 ** \\
        \bottomrule
        \end{tabular}
        \caption{Summary of coefficients from the logistic model}
        \label{tab:log_model}
    \end{minipage}
\end{figure}

\subsection{Model Evaluation}
    To evaluate the model's performance, several statistical tests were used:

    \begin{itemize}
        \item \textbf{Likelihood Ratio Test:} The full model fits significantly better than the null model, indicating that the included variables have a significant effect on predicting the dependent variable.
        \item \textbf{Hosmer-Lemeshow Test:} This test evaluates whether the probabilities predicted by the logistic model differ significantly from those observed in different groups. The results suggest that the model could improve its specification.
        \item \textbf{Area Under the ROC Curve (AUC-ROC):} This provides a visual and quantitative measure of the model's ability to distinguish between classes. An area close to 1 indicates excellent predictive power.
    \end{itemize}

\subsection{Robustness Tests}
    Additional tests were conducted to evaluate the robustness of the results, including:

    \begin{itemize}
        \item \textbf{Multicollinearity:} Checking for potential multicollinearity problems between independent variables.
        \item \textbf{Sensitivity Analysis:} Performing sensitivity analyses to ensure the stability of the estimated coefficients.
    \end{itemize}

\subsection{Coefficient Analysis}

The analysis of the coefficients from the logistic regression model reveals several significant relationships between the independent variables and workplace autonomy in the choice of work schedules. The coefficient for labor income (X1) is negative but extremely small (\( \beta = -0.0000000924 \), \( p < 0.0001 \)), indicating a nearly negligible impact. In contrast, working in medium-sized (\( \beta = 0.4844 \), \( p < 0.0001 \)) and large (\( \beta = 0.1721 \), \( p = 0.0006 \)) companies is associated with greater autonomy, suggesting that these companies offer more flexibility compared to smaller ones.

Tenure in the company (X3) has a negative coefficient (\( \beta = -0.0023 \), \( p < 0.0001 \)), indicating that longer tenure is associated with less autonomy, possibly due to more rigid organizational structures. Regarding workplace location (X4), working in kiosks (\( \beta = 0.7143 \), \( p < 0.0001 \)), rural areas (\( \beta = 0.6548 \), \( p < 0.0001 \)), construction sites (\( \beta = 0.5193 \), \( p < 0.0001 \)) and offices (\( \beta = 0.9899 \), \( p < 0.0001 \)) is associated with greater autonomy. However, working in a vehicle (\( \beta = -0.1524 \), \( p = 0.1832 \)) does not show a significant relationship.

\begin{figure}[h]
    \centering
    \includegraphics[width=0.7\textwidth]{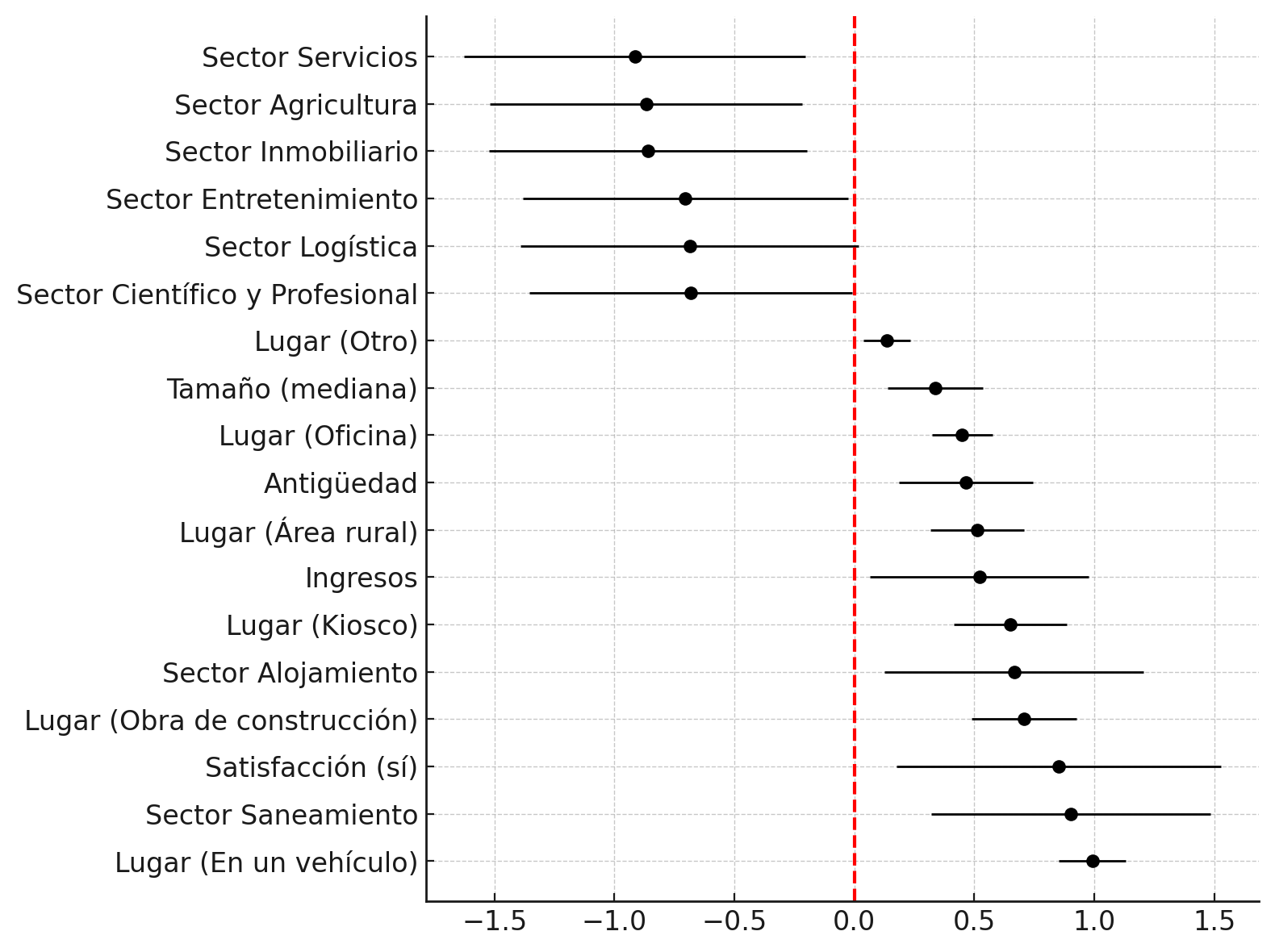}
    \caption{Logistic regression coefficients for workplace autonomy in the choice of work schedules.}
    \label{fig:output-2}
\end{figure}

The coefficient plot (\autoref{fig:output-2}) provides a clear visual representation of these results. In summary, factors such as company size and workplace location significantly influence workplace autonomy, offering a solid foundation for future research and the development of labor policies that promote flexibility and autonomy.

\vspace{-5pt} 
\section{Discussion}

The results of this study highlight the importance of various factors in determining workplace autonomy in the choice of schedules, as explored in the theoretical framework. Autonomy in work schedules, understood as the ability of employees to decide their own working times, is a crucial aspect of the modern work environment that can significantly influence job satisfaction and performance.

One of the most significant findings is the positive relationship between labor income and autonomy in the choice of schedules. Employees with higher incomes tend to have more autonomy, which may be related to their greater bargaining power and status within the company. This result is consistent with previous studies that suggest that employees in higher positions within the organizational hierarchy tend to have more control over their work schedules (Breaugh, 1985; Lopes, Lagoa, and Calapez, 2014).

Tenure in the company is also identified as a positive factor influencing autonomy. Employees with longer tenure tend to have more autonomy in choosing their schedules, which could reflect the trust and value the organization places on them. This relationship may be due to the accumulation of experience and knowledge, allowing the worker to perform tasks more independently. This finding aligns with Karasek's (1979) demand-control theory, which posits that the combination of high autonomy and low job pressure improves employee well-being.

Job satisfaction emerges as a positive and significant predictor of autonomy in schedule choice. Employees who are satisfied with their jobs tend to have more autonomy, suggesting that a positive and supportive work environment can foster flexibility. This finding highlights the importance of organizational policies and practices that promote employee satisfaction to improve not only their well-being but also their perception of autonomy. Studies such as Schwalbe (1985) and Lopes, Lagoa, and Calapez (2014) have documented that workplace autonomy is positively associated with job satisfaction and worker well-being.

Regarding company size, a negative relationship is observed between autonomy in schedule choice and medium-sized companies. This suggests that medium-sized organizations may have more rigid structures and policies compared to small or large companies, limiting employee flexibility. Medium-sized companies may face specific challenges in managing schedules due to the need to maintain a formal structure while seeking to grow and expand. This phenomenon can be explained by organizational decentralization and delegation of authority, which are usually more limited in medium-sized companies (Friedman, 1999).

Workplace location is another relevant factor in determining autonomy. Employees who work in offices, rural areas, kiosks, construction sites, or vehicles have more autonomy in choosing their schedules. This may be due to the nature of the work in these places, which allows for greater flexibility compared to more traditional or structured environments. The variability in working conditions and the need to adapt to different contexts may explain this greater autonomy.

Economic sectors also show significant variations in workplace autonomy. Sectors such as services, agriculture, real estate, entertainment, logistics, and scientific and professional services show less autonomy compared to other sectors. This could reflect the operational particularities and specific demands of these sectors, which may require stricter and less flexible schedules. In contrast, the accommodation and sanitation sectors show greater autonomy, possibly due to the need to adapt to variable demands and the nature of the services offered.

In summary, the findings of this study provide a deep understanding of the factors that influence workplace autonomy in the choice of schedules. Identifying these determinants can guide the design of labor policies and practices that promote autonomy and, in turn, improve employee satisfaction and performance. Future research could expand on these findings by considering other possible factors and applying longitudinal studies to observe how these relationships change over time. Integrating this knowledge into human resource management and organizational design can significantly contribute to creating more flexible and satisfying work environments.

\section{Conclusions}

This study has identified several key factors that influence workplace autonomy in the choice of schedules. The results show that higher labor income and working in larger companies are associated with greater autonomy, while longer tenure in the company tends to reduce it. Additionally, the nature and location of the job play a crucial role in determining autonomy, with locations such as kiosks, rural areas, and offices showing greater flexibility in schedules.

Job satisfaction, while expected as a positive predictor of autonomy, showed a negative relationship in this study, suggesting the need to further investigate other factors that may be influencing this dynamic.

Overall, these findings highlight the importance of considering multiple dimensions of the work environment when designing policies that promote flexibility and autonomy. Fostering an environment that allows employees to decide on their schedules can significantly improve their job satisfaction and performance. Future research could expand on these results by considering other factors and applying longitudinal studies to observe how these relationships evolve over time.


\end{document}